\documentclass{article}
\usepackage{amsmath}
\usepackage{cite}
\usepackage{graphicx}
\usepackage{dcolumn}

\begin{document}

\date{}
\title{On the Raleigh-Ritz variational method. Non-orthogonal basis set}
\author{Francisco M. Fern\'{a}ndez\thanks{%
fernande@quimica.unlp.edu.ar} \\
%EndAName
INIFTA, DQT, Sucursal 4, C. C. 16, \\
1900 La Plata, Argentina}
\maketitle

\begin{abstract}
We overview the main equations of the Rayleigh-Ritz variational method and
discuss their connection with the problem of simultaneous diagonalization of
two Hermitian matrices.
\end{abstract}

\section{Introduction}

\label{sec:intro}

The Rayleigh-Ritz variational method (RR) is one of the approximate methods
most commonly used in the study of the electronic structure of atoms and
molecules\cite{P68,SO96}. One of its main advantages is that it provides
increasingly accurate upper bounds to all the eigenvalues of the Hamiltonian
operator of the system\cite{M33,F22}. In this paper we provide a
comprehensible overview of the approach and illustrate some of its relevant
points by means of a simple problem.

\section{The Rayleigh-Ritz variational method}

\label{sec:RR}

The starting point of our analysis is a linearly independent set of vectors $%
\mathcal{V}=$ $\left\{ f_{1},f_{2},\ldots \right\} $. Clearly, the only
solution to the vector equation
\begin{equation}
\sum_{i=1}^{N}a_{i}f_{i}=0,  \label{eq:lin_indep}
\end{equation}
is $a_{i}=0$ for all $i=1,2,\ldots ,N$. If we apply the bras $\left\langle
f_{j}\right| $, $j=1,2,\ldots ,N$, to this equation from the left we obtain
\begin{equation}
\sum_{i=1}^{N}S_{ji}a_{i}=0,\;j=1,2,\ldots ,N,  \label{eq:lin_indep_2}
\end{equation}
where $S_{ij}=\left\langle f_{i}\right| \left. f_{j}\right\rangle $. We have
an homogeneous system of $N$ linear equations with $N$ unknowns $a_{i}$ with
the only solution $a_{i}=0$. Consequently, $\left| \mathbf{S}\right| \neq 0$
where $\mathbf{S}=\left( S_{ij}\right) _{i,j=1}^{N}$ is an $N\times N$
Hermitian matrix and $\left| ...\right| $ stands for determinant. Note that $%
S_{ij}=S_{ji}^{*}$ so that $\mathbf{S}^{\dagger }=\mathbf{S}$ where $\dagger
$ stands for adjoint. The matrix $\mathbf{S}$ is commonly called overlap
matrix\cite{P68}.

Let $\mathbf{v}$ be an eigenvector of $\mathbf{S}$ with eigenvalue $s$, $%
\mathbf{Sv}=s\mathbf{v}$, then $\mathbf{v}^{\dagger }\mathbf{Sv}=s\mathbf{v}%
^{\dagger }\mathbf{v}$. If $v_{i}$, $i=1,2,\ldots ,N$, are the elements of
the $N\times 1$ column vector $\mathbf{v}$ then
\begin{equation}
\mathbf{v}^{\dagger }\mathbf{Sv}=\left\langle
\sum_{i=1}^{N}v_{i}f_{i}\right| \left. \sum_{j=1}^{N}v_{j}f_{j}\right\rangle
>0,  \label{eq:v+Sv}
\end{equation}
and we conclude that $s>0.$ In other words, the overlap matrix $\mathbf{S}$
is positive definite.

We are interested in the eigenvalue equation
\begin{eqnarray}
H\psi _{n} &=&E_{n}\psi _{n},\;n=1,2,\ldots ,  \nonumber \\
E_{1} &\leq &E_{2}\leq \ldots ,\;\left\langle \psi _{i}\right. \left| \psi
_{j}\right\rangle =\delta _{ij},  \label{eq:eigenvalue_eq_H}
\end{eqnarray}
for an Hermitian operator $H$. In order to solve it approximately we propose
and ansatz of the form
\begin{equation}
\varphi =\sum_{j=1}^{N}c_{j}f_{j},  \label{eq:varphi}
\end{equation}
where $\mathcal{V}=\left\{ f_{1},f_{2},\ldots \right\} $ is not only assumed
to be linearly independent but also complete.

The RR variational method consists of minimizing the integral
\begin{equation}
W=\frac{\left\langle \varphi \right| H\left| \varphi \right\rangle }{%
\left\langle \varphi \right| \left. \varphi \right\rangle },  \label{eq:W}
\end{equation}
with respect to the expansion coefficients $c_{j}$
\begin{equation}
\frac{\partial W}{\partial c_{j}}=0,\;j=1,2,\ldots ,N\text{\ .}
\label{eq:dW/dcj}
\end{equation}
This equation leads to the so-called secular equation\cite{P68,SO96}
\begin{equation}
\sum_{j=1}^{N}\left( H_{ij}-WS_{ij}\right) c_{j}=0,\;i=1,2,\ldots ,N,
\label{eq:secular_eq}
\end{equation}
where, $H_{ij}=\left\langle f_{i}\right| H\left| f_{j}\right\rangle $. There
are nontrivial solutions $c_{j}$, $j=1,2,\ldots ,N$, provided that the
secular determinant vanishes
\begin{equation}
\left| \mathbf{H}-W\mathbf{S}\right| =0,  \label{eq:sec_det}
\end{equation}
where $\mathbf{H}=\left( H_{ij}\right) _{i,j=1}^{N}$ is an $N\times N$
Hermitian matrix.

For each of the roots of the secular determinant (\ref{eq:sec_det}), $%
W_{1}\leq W_{2}\leq \ldots \leq W_{N}$, we derive an approximate solution;
for example, when $W=W_{k}$ we have
\begin{equation}
\varphi _{k}=\sum_{j=1}^{N}c_{jk}f_{j},  \label{eq:varphi_k}
\end{equation}
and the secular equation (\ref{eq:secular_eq}) can be rewritten
\begin{equation}
\sum_{j=1}^{N}H_{ij}c_{jk}=\sum_{j=1}^{N}W_{k}S_{ij}c_{jk}=\sum_{j=1}^{N}%
\sum_{m=1}^{N}S_{ij}W_{m}\delta _{mk}c_{jm}.  \label{eq:secular_eq_2}
\end{equation}
If we define the $N\times N$ matrices $\mathbf{W}=\left( W_{i}\delta
_{ij}\right) _{i,j=1}^{N}$ and $\mathbf{C}=\left( c_{ij}\right) _{i,j=1}^{N}$%
then this equation can be rewritten in matrix form as
\begin{equation}
\mathbf{HC}=\mathbf{SCW,}  \label{eq:secular_eq_matrix}
\end{equation}
which is equivalent to
\begin{equation}
\mathbf{C}^{-1}\mathbf{S}^{-1}\mathbf{HC}=\mathbf{W,}
\label{eq:secular_eq_matrix_2}
\end{equation}
and the procedure reduces to the diagonalization of the matrix $\mathbf{S}%
^{-1}\mathbf{H}$ by means of the invertible matrix $\mathbf{C}$. Note that $%
\mathbf{S}^{-1}$ exists because $\mathbf{S}$ is positive definite as argued
above.

In order to determine the coefficients $c_{jk}$ completely, we require that $%
\left\langle \varphi _{i}\right. \left| \varphi _{j}\right\rangle =\delta
_{ij}$ that leads to
\begin{equation}
\left\langle \varphi _{i}\right. \left| \varphi _{j}\right\rangle
=\sum_{k=1}^{N}\sum_{m=1}^{N}c_{ki}^{*}c_{mj}\left\langle f_{k}\right|
\left. f_{m}\right\rangle =\delta _{ij},
\end{equation}
that in matrix form reads
\begin{equation}
\mathbf{C}^{\dagger }\mathbf{SC}=\mathbf{I,}  \label{eq:C+SC}
\end{equation}
where $\mathbf{I}$ is the $N\times N$ identity matrix. It follows from
equations (\ref{eq:C+SC}) and (\ref{eq:secular_eq_matrix}) that
\begin{equation}
\mathbf{C}^{\dagger }\mathbf{HC}=\mathbf{W.}  \label{eq:C+HC}
\end{equation}
It is clear that there exists an invertible matrix ($\mathbf{C}$) that
transforms two Hermitian matrices ($\mathbf{H}$ and $\mathbf{S}$), one of
them positive definite ($\mathbf{S}$), into diagonal form. This procedure is
well known in the mathematical literature\cite{BC84}. However, it is most
important to note that equations (\ref{eq:C+SC}) and (\ref{eq:C+HC}) are not
what we commonly know as matrix diagonalization. In fact, the eigenvalues of
$\mathbf{S}$ are not unity and the eigenvalues of $\mathbf{H}$ are not the
RR eigenvalues $W_{i}$. We will illustrate this point in section~\ref
{sec:example} by means of a simple example. It is also worth noting that
that we cannot obtain $\mathbf{C}$ neither from (\ref{eq:C+SC}) or (\ref
{eq:C+HC}). One obtains the matrix $\mathbf{C}$ in the process of
diagonalizing $\mathbf{S}^{-1}\mathbf{H}$ as in equation (\ref
{eq:secular_eq_matrix_2}) and the remaining undefined matrix elements $%
c_{ij} $ from equation (\ref{eq:C+SC}).

Since $\mathbf{S}$ is positive definite, we can define $\mathbf{S}^{1/2}$.
The matrix $\mathbf{U}=\mathbf{S}^{1/2}\mathbf{C}$ is unitary as shown by
\begin{equation}
\mathbf{U}^{\dagger }\mathbf{U=C}^{\dagger }\mathbf{\mathbf{S}^{1/2}S}^{1/2}%
\mathbf{C}=\mathbf{I.}  \label{eq:U+U}
\end{equation}
On substituting $\mathbf{C}=\mathbf{S}^{-1/2}\mathbf{U}$ into equation (\ref
{eq:C+HC}) we obtain
\begin{equation}
\mathbf{U}^{\dagger }\mathbf{S}^{-1/2}\mathbf{HS}^{-1/2}\mathbf{U}=\mathbf{W.%
}  \label{eq:U+SHSU}
\end{equation}
This equation is just the standard diagonalization of the Hermitian matrix $%
\mathbf{\mathbf{S}^{-1/2}HS}^{-1/2}$.

If the basis set is orthonormal, $\left\langle f_{i}\right| \left.
f_{j}\right\rangle =\delta _{ij}$, then $\mathbf{S}=\mathbf{I}$, $\mathbf{C}%
^{\dagger }=\mathbf{C}^{-1}$ and the secular equation (\ref
{eq:secular_eq_matrix_2}) becomes
\begin{equation}
\mathbf{C}^{\dagger }\mathbf{HC}=\mathbf{W.}  \label{eq:sec_eq_ortho}
\end{equation}
In this particular case, the eigenvalues of the matrix $\mathbf{H}$ are the
RR eigenvalues $W_{i}$. Note that equations (\ref{eq:C+HC}) and (\ref
{eq:sec_eq_ortho}) look identical but were derived under different
assumptions (they agree only when $\mathbf{S}=\mathbf{I}$).

\section{Simple example}

\label{sec:example}

As a simple example we consider the dimensionless eigenvalue equation
\begin{equation}
H\psi =E\psi ,\;H=-\frac{1}{2}\frac{d^{2}}{dx^{2}}+\lambda x,\;\psi (0)=\psi
(1)=0.  \label{eq:Schro_example}
\end{equation}
In order to illustrate the RR variational method with a non-orthogonal basis
set we choose $f_{i}(x)=x^{i}(1-x)$, $i=1,2,\ldots $, that satisfy the
boundary conditions at $x=0$ and $x=1$.

A straightforward calculation shows that
\begin{equation}
S_{ij}=\frac{2}{\left( i+j+1\right) \left( i+j+2\right) \left( i+j+3\right) }%
,  \label{eq:S_(ij)_example}
\end{equation}
and
\begin{equation}
H_{ij}=\frac{ij}{\left( i+j\right) \left( i+j+1\right) \left( i+j-1\right) }+%
\frac{2\lambda }{\left( i+j+2\right) \left( i+j+3\right) \left( i+j+4\right)
}.  \label{eq:H_(ij)_example}
\end{equation}
Tables \ref{tab:RRNORT0} and \ref{tab:RRNORT1} show the RR eigenvalues $%
W_{i} $, $i=1,2,3,4$, for $\lambda =0$ and $\lambda =1$, respectively. We
appreciate that the approximate eigenvalues converge from above as expected%
\cite{M33,F22}.

In what follows, we illustrate some of the general results of section~\ref
{sec:RR} for the simplest case $N=2$ when $\lambda =0$. The matrices are
\begin{equation}
\mathbf{S}=\frac{1}{60}\left(
\begin{array}{ll}
2 & 1 \\
1 & \frac{4}{7}
\end{array}
\right) ,\;\mathbf{H}=\frac{1}{12}\left(
\begin{array}{ll}
2 & 1 \\
1 & \frac{4}{5}
\end{array}
\right) ,
\end{equation}
and we obtain
\begin{equation}
\mathbf{C}^{-1}\mathbf{S}^{-1}\mathbf{HC}=\mathbf{W}=\left(
\begin{array}{ll}
5 & 0 \\
0 & 21
\end{array}
\right) ,\;\mathbf{C}=\sqrt{30}\left(
\begin{array}{ll}
1 & \sqrt{7} \\
0 & -2\sqrt{7}
\end{array}
\right) .
\end{equation}
One can easily verify that these matrices already satisfy equations (\ref
{eq:C+SC}) and (\ref{eq:C+HC}). On the other hand, the symmetric matrices $%
\mathbf{S}$ and $\mathbf{H}$ can be diagonalized in the usual way by
orthogonal matrices that we call $\mathbf{U}_{S}$ and $\mathbf{U}_{H}$,
respectively.
\begin{eqnarray}
\mathbf{U}_{S}^{\dagger }\mathbf{SU}_{S} &=&\frac{1}{420}\left(
\begin{array}{ll}
9-\sqrt{74} & 0 \\
0 & 9+\sqrt{74}
\end{array}
\right) ,\;  \nonumber \\
\mathbf{U}_{S} &=&\left(
\begin{array}{ll}
\sqrt{\frac{1}{2}-\frac{5\sqrt{174}}{148}} & \sqrt{\frac{1}{2}+\frac{5\sqrt{%
174}}{148}} \\
-\sqrt{\frac{1}{2}+\frac{5\sqrt{174}}{148}} & \sqrt{\frac{1}{2}-\frac{5\sqrt{%
174}}{148}}
\end{array}
\right) ,  \nonumber \\
\mathbf{U}_{H}^{\dagger }\mathbf{HU}_{H} &=&\frac{1}{60}\left(
\begin{array}{ll}
7-\sqrt{34} & 0 \\
0 & 70+\sqrt{34}
\end{array}
\right) ,\;  \nonumber \\
\mathbf{U}_{H} &=&\left(
\begin{array}{ll}
\sqrt{\frac{1}{2}-\frac{3\sqrt{34}}{68}} & \sqrt{\frac{1}{2}+\frac{3\sqrt{34}%
}{68}} \\
-\sqrt{\frac{1}{2}+\frac{3\sqrt{34}}{68}} & \sqrt{\frac{1}{2}-\frac{3\sqrt{34%
}}{68}}
\end{array}
\right)  \label{eq:USSUS,UHHUH}
\end{eqnarray}
We clearly see that the eigenvalues of $\mathbf{S}$ are not unity and those
of $\mathbf{H}$ are not the RR eigenvalues $W_{i}$ as argued in section~\ref
{sec:RR}.

Using equation (\ref{eq:USSUS,UHHUH}) one can easily obtain
\begin{equation}
\mathbf{S}^{1/2}=\left(
\begin{array}{ll}
\sqrt{\frac{233}{8880}+\frac{7\sqrt{7}}{8880}} & \sqrt{\frac{21}{2960}-\frac{%
7\sqrt{7}}{8880}} \\
\sqrt{\frac{21}{2960}-\frac{7\sqrt{7}}{8880}} & \sqrt{\frac{151}{62160}+%
\frac{7\sqrt{7}}{8880}}
\end{array}
\right) .
\end{equation}

\section{Conclusions}

\label{sec:conclusions}

We have shown that the main equations of the Rayleigh-Ritz variational method%
\cite{P68,SO96} lead to the mathematical problem of diagonalization of two
Hermitian matrices\cite{BC84}. Although equations (\ref{eq:C+SC}) and (\ref
{eq:C+HC}) are discussed in some textbooks on quantum chemistry, the latter
does not appear to be correctly interpreted\cite{P68}.

\begin{table}[tbp]
\caption{Convergence of the Rayleigh-Ritz variational method with a
non-orthogonal basis set for $\lambda=0$}
\label{tab:RRNORT0}
\begin{center}
\par
\begin{tabular}{D{.}{.}{3}D{.}{.}{11}D{.}{.}{11}D{.}{.}{11}D{.}{.}{11}}
\hline \multicolumn{1}{l}{$N$}&\multicolumn{1}{c}{$E_1$}&
\multicolumn{1}{c}{$E_2$} &
\multicolumn{1}{c}{$E_3$} & \multicolumn{1}{c}{$E_4$} \\
\hline
  4&   4.934874810 &  19.75077640 &  51.06512518 &  100.2492235  \\
  5&   4.934802217 &  19.75077640 &  44.58681182 &  100.2492235  \\
  6&   4.934802217 &  19.73923669 &  44.58681182 &  79.99595777  \\
  7&   4.934802200 &  19.73923669 &  44.41473408 &  79.99595777  \\
  8&   4.934802200 &  19.73920882 &  44.41473408 &  78.97848206  \\
  9&   4.934802200 &  19.73920882 &  44.41322468 &  78.97848206  \\
 10&   4.934802200 &  19.73920880 &  44.41322468 &  78.95700917  \\
 11&   4.934802200 &  19.73920880 &  44.41321981 &  78.95700917  \\
 12&   4.934802200 &  19.73920880 &  44.41321981 &  78.95683586  \\
 13&   4.934802200 &  19.73920880 &  44.41321980 &  78.95683586  \\
 14&   4.934802200 &  19.73920880 &  44.41321980 &  78.95683521  \\
 15&   4.934802200 &  19.73920880 &  44.41321980 &  78.95683521  \\
 16&   4.934802200 &  19.73920880 &  44.41321980 &  78.95683520  \\
 17&   4.934802200 &  19.73920880 &  44.41321980 &  78.95683520  \\
 18&   4.934802200 &  19.73920880 &  44.41321980 &  78.95683520  \\
 19&   4.934802200 &  19.73920880 &  44.41321980 &  78.95683520  \\
 20&   4.934802200 &  19.73920880 &  44.41321980 &  78.95683520  \\

 \end{tabular}
\par
\end{center}
\end{table}

\begin{table}[tbp]
\caption{Convergence of the Rayleigh-Ritz variational method with a
non-orthogonal basis set for $\lambda=1$}
\label{tab:RRNORT1}
\begin{center}
\par
\begin{tabular}{D{.}{.}{3}D{.}{.}{11}D{.}{.}{11}D{.}{.}{11}D{.}{.}{11}}
\hline \multicolumn{1}{l}{$N$}&\multicolumn{1}{c}{$E_1$}&
\multicolumn{1}{c}{$E_2$} &
\multicolumn{1}{c}{$E_3$} & \multicolumn{1}{c}{$E_4$} \\
\hline
   4 & 5.432678349 & 20.25175971 & 51.56499993 &  100.7505620   \\
   5 & 5.432608286 & 20.25141191 & 45.08766430 &  100.7488422  \\
   6 & 5.432607868 & 20.23989706 & 45.08714181 &  80.49674963  \\
   7 & 5.432607855 & 20.23989074 & 44.91514957 &  80.49606992  \\
   8 & 5.432607855 & 20.23986309 & 44.91512224 &  79.47878520   \\
   9 & 5.432607855 & 20.23986306 & 44.91361487 &  79.47871372  \\
  10 & 5.432607855 & 20.23986304 & 44.91361453 &  79.45724985  \\
  11 & 5.432607855 & 20.23986304 & 44.91360967 &  79.45724783  \\
  12 & 5.432607855 & 20.23986304 & 44.91360967 &  79.45707467  \\
  13 & 5.432607855 & 20.23986304 & 44.91360966 &  79.45707465  \\
  14 & 5.432607855 & 20.23986304 & 44.91360966 &  79.45707400   \\
  15 & 5.432607855 & 20.23986304 & 44.91360966 &  79.45707400   \\
  16 & 5.432607855 & 20.23986304 & 44.91360966 &  79.45707400   \\
  17 & 5.432607855 & 20.23986304 & 44.91360966 &  79.45707400   \\
  18 & 5.432607855 & 20.23986304 & 44.91360966 &  79.45707400   \\
  19 & 5.432607855 & 20.23986304 & 44.91360966 &  79.45707400   \\
  20 & 5.432607855 & 20.23986304 & 44.91360966 &  79.45707400   \\

 \end{tabular}
\par
\end{center}
\end{table}

\end{document}